# E- Learning: An effective pedagogical tool for learning

Boumedyen [#1], Kaneez [*2], Rafael [#3] Victor [**4]
[#1]Information System, University of Nizwa, [#3]Information System, SPIRAN University
[#1]Birkut- ul- Mauz,Nizwa, Sultanate of Oman, , [#3]St. Petersburg, Russia
[*2] Management, University of Nizwa
Birkut- ul- Mauz,Nizwa, Sultanate of Oman
, [**4]Information System, SPIRAN University,, St. Petersburg, Russia

*Abstract*— In the info-tech age E-Methods of learning are becoming the most important vehicle in disseminating knowledge in higher education institutions. This sector is growing and changing at a rapid speed due to developments in technologies. But teaching is an art. Can there be fun learning with raw and dry technology? How can we make the best use of E- Methods, can we make the required information and data available to the students in a flexible manner, at ease all the time? What are the advantages of traditional methods of teaching and learning? Is E-learning a progressive stage incubating all the benefits of the Manual learning or it is only a window dressing on the face of advancement? Can we convert the boring, tedious subjects into interactive, monotony breaking joyous learning? In this paper the researchers have focused on the modernization of E- Pedagogy vis-à-vis the traditional method of learning. They have highlighted the effectiveness of using the E-learning elements and various E- Methods. This work has used the decision tree algorithms particularly Classifiers.trees.J48 The obtained results show that using online examination attribute plays major role in increasing the average grade of the class in higher education. The novelty of this work is that the researchers have focused on the teaching methodology used by the faculty members and the tools available in the universities. We believe that this work will play a constructive role in building higher education system. Our generated rules/output can be used by the decision makers in the improvement of higher education system processes.

Key words: E- learning, higher education systems, modernization, decision tree algorithm.

## VI. INTRODUCTION

I n the info-tech age E-Methods of learning are becoming the most important vehicle in disseminating knowledge in higher education institutions. This sector is growing and changing at a rapid speed due to developments in technologies.

But teaching is an art. Ever since Socrates thought of teaching geometry to the slave boy in Plato's Meno, the nature of learning has been an active topic of investigation [1]. Both undergraduate and graduate courses are experiencing a migration away from the traditional classroom and toward a greater emphasis for electronic delivery of content [2]. This trend is equally applicable on all departments and schools in the university system but is especially critical in business schools, since the preparation of students for successful business careers will depend on the students' abilities to accurately assess the quality of teaching and rapidly adapt to the changing pedagogy that reflect radical technological advances. The researchers have tried to examine the adoption of pedagogical changes in Higher Education with respect to the introduction and growth of e-learning. Our professed aim is to use e-learning to improve the quality of the teaching-learning experience for faculties and students. Unfortunately, the teaching quality ranks poor in relation to most of changes required in higher education. Out of the many traditional and non traditional pedagogical tools for learning such as oral conversation, case study method, group discussions, classroom games, simulation exercises, distance learning etc E-learning emerges to be the best one. This is high time when higher education institutions think seriously on improving faculty-student learning, and one humble step in this direction is to exploit e-learning. There is always resistance to change, each time we have to ride on a new wave of technological innovation. The novelty of this work is that unlike other researches which focuses on students learning behavior we have focused on the teaching pedagogy used by the faculty members and the instruments available in the universities. Usually the blame goes to the students for their non-performance as being inattentive or careless or indolent, lethargic or idle but we ignore the important vehicle in education that is the Instructor, education imparting pedagogy, electronic media, and instruments of teaching and available resources. We believe that this work will play a constructive role in building higher education system. It will encourage the use of technology in teaching. Our generated rules/output can be used by the decision makers in the improvement of higher education system processes. In this paper we have studied the instructor's behavior and the class attributes. Further research can be carried out on the basis of this work so as to compare the student's behavior or learning attitude with the present study.

What is M-learning (Manual learning)?

The classroom face to face teaching with the help of text books and scheduling the class timings, following strict timetable, physical presence, hardcopy of class notes, books, assignments, pre-decided meeting places and timings, face to face interaction, communication and question answer sessions, group discussions and physical participation in educational games, usually one teacher and students of similar age form traditional learning [1] [2].

*C. What is E- learning?*

Using the electronic methods as learning tools such as multimedia, internet, computer, software, online textual materials to streaming video to chat rooms and discussion boards which means that students have many more choices in an E-learning environment than they had in a more M-learning, face-to-face environment. Catalog of media-enhanced Power-Point slides, streaming video lectures, some interactive Excel-based practice problems which may be individualized to the students form part of E-learning [3]

*D. Why is e-learning important for HE (Higher Education)?*

[3]Higher education is the link between knowledge gained and practical implementation in industry. E-learning allows students access to learning without the constraints of time and location , [4] Some of the added benefits of E-education include flexibility; ease of participation; absence of labeling due to such things as race, gender, and appearance; training in electronic communication; and exposing students to information technology [5] E-learning in general and online college education specifically are having a profound effect on the future of postsecondary education and is transforming the educational model from an instructor-driven to an interactive and community-driven educational environment in which all students share responsibility for learning outcomes [6].

VII. RELATED WORK

Recently many researchers have worked to enhance and evaluate the higher education tasks. Some researchers have proposed methods and architectures by using data mining in higher education. In [7] they investigated the current trends in improving the higher education systems, to understand from the outside which factors might create devoted students. They used Data mining methods to extract valuable information from existing students so as to handle prospective students in a better manner. They have generated some rules which may be used to understand the behavioral pattern and learning attitude of the prospective enrollments at an early stage and thus the concentration of effort in higher education systems may be implemented. The research by [8] proposed a model to represent how data mining is used in higher educational system to improve the efficiency and effectiveness of the traditional processes. [9] Discusses different AI technologies and compares them with genetic algorithm based induction of decision trees and discusses why the approach has a potential for developing an alert tool. The researchers in [10] proposed a model to represent how data mining is used in Institutes of higher learning to improve the competence and effectiveness of the traditional processes. In this model a guideline was presented for higher educational system to improve their decision-making processes. The Work by [11] is to use Rough Set theory as classification approach to analyze student data. In this research the Rosetta toolkit has been used to evaluate the student data to describe the dependencies between the attributes and the student status.

Many other related Information and works can be found in [12,13,14,15,16,17,18,19,20,21,22,23,24].

VIII. DATA COLLECTION

For implementing this work we have used the data base of specific university. For some reason we are not able to reveal the name of the university. The data base has information about the teaching attributes and overall marks achieved by the class for one semester. The database has information about marks of 139 courses from different domains (subjects) for one semester, provided by specific university. Each of the attributes in this data set has dichotomous values- Yes and No value. If the instructor uses the particular teaching method the value will be equal to 'Yes' and if not the value will be 'No'. For simplification of the processing of the collected data set we replaced the nominal value to numerical value i.e. Yes=1, and No=0. All the attributes have two instances except the 'mark' attribute which has six instances i.e. D, D+, C, C+, B, B+ (see fig.3.1)

Fig 3.1 Data Set Sample in binary view

IX. EXPERIMENT

Using machine learning to find the conditions that are suitable for improving higher education system using E- learning methods and M- learning (manual learning) methods.

The general instances in the data set are characterized by the values of attributes that measures different aspects of the instances. In this work there are 25 attributes such as Using power point, Using multimedia in the class, Using physical models, Using e-mail for communication with students, Using websites to display instruction material, Narrative discussions in classroom, Using white board-marker for teaching, Using mobile for communication, Using group discussion in the class, Using computers in the lab, Using any kind of software, Using the internet for giving assignments, Using internet for submitting assignments, Using local Eduwave for teaching, Physical Visit to instructor's office, Using online system for registration, Using Online examination, Displaying marks on line, Using physical handouts, Using physical instruments, Using facial expressions, movement of hands, Contacting instructor by mail, Teaching with the help of book, Student presents their work using PowerPoint and Marks. The outcome shall be the effect on student's performance due to the use of E-learning, M- learning or any other teaching method and to highlight which attributes play the main role in improving the student's performance. In its simplest form as shown in fig 3.1, we have renamed the attributes for simplifying the process of the data set as shown in fig 4.1. For Example we have assigned label 'UOE' to describe 'Using online exam' and so on.

Figure 4.1 Renamed attributes of data set

We prepared description for each attribute as follows- Name, Type, Missing, Distinct and Unique; see the following for further detail.
Sample Selected Attributes:

Using Power Point

| Name: Using power point | | Type: Numeric |
| Missing: 0 (0%) | Distinct: 2 | Unique: 0 (0%) |
| Statistic | Value | |
| Minimum | 0 | |
| Maximum | 1 | |
| Mean | 0.777 | |
| StdDev | 0.418 | |

Figure 4.2 Value description of using power point attribute

From the above figure 4.2 the Using power Point attribute has no missing values ,Two distinct values and no unique values .

| Name: Using mulimedia in the class | | Type: Numeric |
| Missing: 0 (0%) | Distinct: 2 | Unique: 0 (0%) |
| Statistic | Value | |
| Minimum | 0 | |
| Maximum | 1 | |
| Mean | 0.245 | |
| StdDev | 0.431 | |

Figure 4.3 Value description of using multimedia in the class attribute

| Name: using physical models | | Type: Numeric |
| Missing: 0 (0%) | Distinct: 2 | Unique: 0 (0%) |
| Statistic | Value | |
| Minimum | 0 | |
| Maximum | 1 | |
| Mean | 0.712 | |
| StdDev | 0.454 | |

Figure 4.4 Value description of using physical models attribute

Figure 4.5 Value description for using facial expression attribute

Figure 4.5 Value description for using mark attribute

Figure 4.6 displays Histogram and it shows how often each of six values of class, Mark, occurs for each values of different attributes.

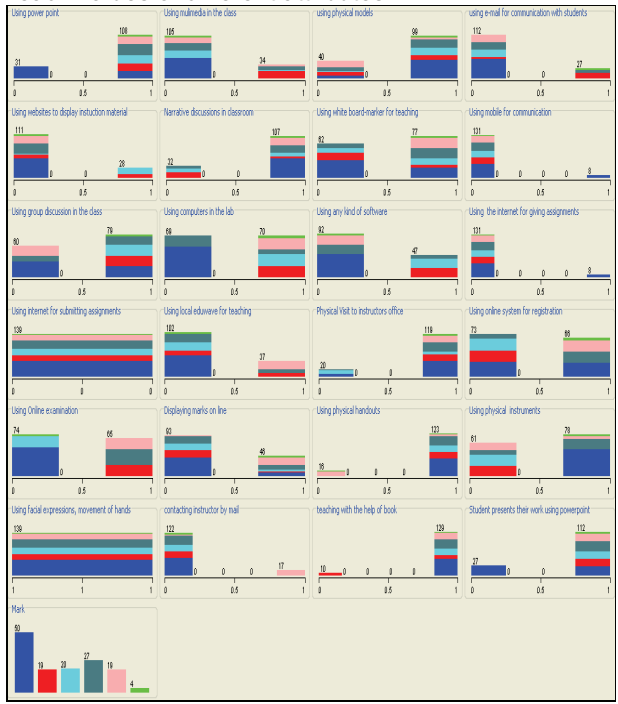

Figure 4.6 Histogram showing how often each of six values of class, Mark, occurs for each values of different attributes.

The histogram shows the distribution of the class as a function of these attributes.
We used the J4.8 algorithm to implement The C4.5 decision tree.

TABLE I
Result-

Run information :
Scheme:      .classifiers.trees.J48 -C 0.25 -M 2
Relation:     Survey on E- Learning
Instances:   139
Attributes:   25
          Using power point
          Using multimedia in the class
          Using physical models
          Using e-mail for communication with students
          Using websites to display instruction material
          Narrative discussions in classroom
          Using white board-marker for teaching
          Using mobile for communication
          Using group discussion in the class
          Using computers in the lab
          Using any kind of software
          Using the internet for giving assignments
          Using internet for submitting assignments
          Using local Eduwave for teaching
          Physical Visit to instructor's office
          Using online system for registration
          Using Online examination
          Displaying marks on line
          Using physical handouts
          Using physical instruments
          Using facial expressions, movement of hands
          Contacting instructor by mail
          Teaching with the help of book
          Student presents their work using PowerPoint
          Mark
Test mode:   10-fold cross-validation
Classifier model (full training set)
J48 pruned tree:
  Using Online examination <= 0
  Using computers in the lab <= 0: D (50.0)
  Using computers in the lab > 0
  Using any kind of software <= 0: D+ (4.0)
  Using any kind of software > 0: B (20.0)
  Using online examination > 0
  Contacting instructor by mail <= 0
  Using any kind of software <= 0
  Using multimedia in the class <= 0: C (19.0)
  Using multimedia in the class > 0: C+ (3.0)
  Using any kind of software > 0

```
  Using physical models <= 0
   Teaching with the help of book <= 0: B+ (6.0)
   Teaching with the help of book > 0: C (11.0/3.0)
   Using physical models > 0: B+ (10.0)
   Contacting instructor by mail > 0: C+ (16.0)

Number of Leaves:           9

Size of the tree:          17
```

Stratified cross-validation   :
Figure 4.7  summarize the experiment details

```
Correctly Classified Instances         131               94.2446 %
Incorrectly Classified Instances         8                5.7554 %
Kappa statistic                          0.9258
Mean absolute error                      0.021
Root mean squared error                  0.1183
Relative absolute error                  8.1055 %
Root relative squared error             32.9231 %
Total Number of Instances              139

=== Detailed Accuracy By Class ===

               TP Rate  FP Rate  Precision  Recall  F-Measure  ROC Area  Class
               1        0        1          1       1          1         D
               0.842    0.033    0.8        0.842   0.821      0.985     B+
               0.95     0        1          0.95    0.974      0.975     B
               0.852    0.027    0.885      0.852   0.868      0.989     C
               1        0        1          1       1          1         C+
               1        0.007    0.8        1       0.889      0.996     D+
Weighted Avg.  0.942    0.01     0.944      0.942   0.943      0.992

=== Confusion Matrix ===

  a  b  c  d  e  f   <-- classified as
 50  0  0  0  0  0 |  a = D
  0 16  0  3  0  0 |  b = B+
  0  0 19  0  0  1 |  c = B
  0  4  0 23  0  0 |  d = C
  0  0  0  0 19  0 |  e = C+
  0  0  0  0  0  4 |  f = D+
```

Figure 4.7 Summary of the result

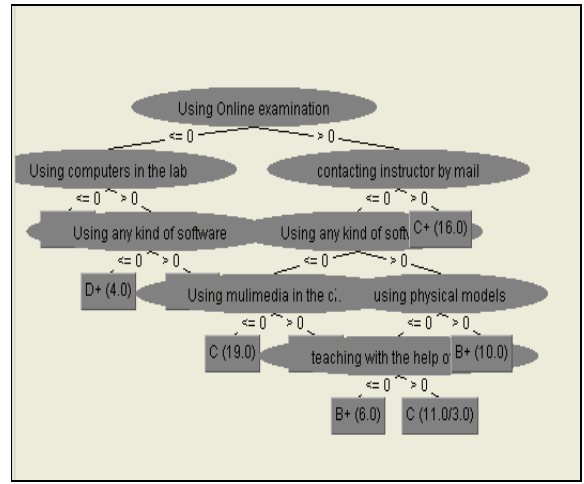

Figure 4.8 Decision Tree for experiment attributes.

To give a better presentation of the decision tree we have used acronyms for each attributes as described if Figure 4.9 .

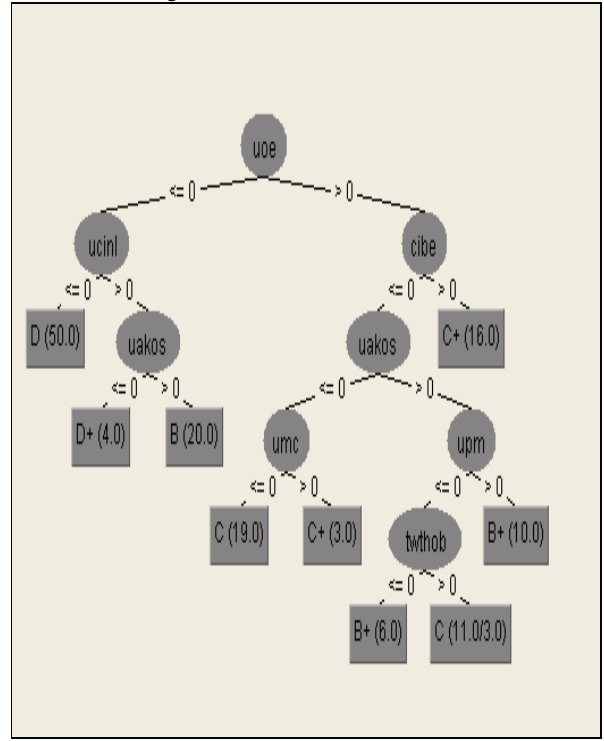

Figure 4.9 Decision tree using acronyms

### X.  OUTCOME/RESULTS

From the decision tree as shown above, we notice that the top node represents the entire data i.e.

139 instances and 28 attributes. The classification tree algorithm finds out that the best way to explain the dependent variable 'Mark' is by using variable 'UOE' which stands for 'Using online Examination'. Using the categories of the variable 'UOE' two different groups (under condition) were observed. If the instructor does not use 'uoe' (using online examination), next question is 'ucinl'(using computer in lab), we can conclude that:

If the instructor neither uses 'uoe'(Online Examination) nor uses 'ucinl' (Using computers in the lab), the 'mark' of the student will be 'D'.

In the same manner if we study other attributes on each level, depending on the decision tree algorithm to trace each path, we generate the following proposals and rules.

Sample of the rule –     >0   = Yes
    <=0   =No

If 'uoe'(using online examination)<=0 'ucinl'(Using computers in the lab) <=0 then the mark is D

If 'uoe'(using online examination)<=0 'ucinl'(Using computers in the lab) >0 and 'uakos'(using any kind of software)<=0 then the mark is D+

If 'uoe'(using online examination)<=0 'ucinl'(Using computers in the lab) >0 and 'uakos'(using any kind of software)>0 then the mark is B

If 'uoe' (using online examination) >0 and 'cibe' (contacting instructor by mail)>0 then the mark is C+

If 'uoe' (using online examination) >0 and 'cibe'(contacting instructor by mail)<=0 and 'uakos'(using any kind of software)<=0 and 'umc'(using multimedia in class)<=0 then the result is C

If 'uoe'(using online examination) >0 and 'cibe'(contacting instructor by mail)<=0 and 'uakos'(using any kind of software)<=0 and 'umc'(using multimedia in class)>0 then the result is C+

If 'uoe' (using online examination) >0 and 'cibe'(contacting instructor by mail)<=0 and 'uakos' (using any kind of software)>0 and 'upm'(using physical models)>0 then the result is B+

If 'uoe'(using online examination) >0 and 'cibe'(contacting instructor by mail)<=0 and 'uakos'(using any kind of software) >0 and 'upm'(using physical models)<=0 and 'twthob'(teaching with the help of books)>0 then the result is C

If 'uoe' (using online examination) >0 and 'cibe'(contacting instructor by mail)<=0 and 'uakos'(using any kind of software) >0 and 'upm'(using physical models)<=0 and 'twthob'(teaching with the help of books)<=0 then the result is B+

XI. CONCLUSION

This study extends the research reported by [18] and enhances our understanding of the relationship between various attributes of teaching in a number of ways. First, our decision tree confirmed that among the various pedagogies, E-learning methods play major role in enhancing the performance of the students. Our findings suggest that the E-learning methods require the consideration of a number of interrelated decisions and antecedent conditions. Successful implementation of E- educational delivery takes a commitment from both the students and the faculty and is not as simple as merely using computer in the labs or contacting the instructor by mail. Second, the study scores the fact that classes that integrate use of software and using computers while teaching even though there is no online examination yet their effect on the marks is significant. Third, the resultant mark of the student, if multimedia is used in the classroom even though no any software is being used and the students communicate with the instructor via mail and appear for on line examination, is better than the students for whom multimedia was not used. Fourthly, the marks of the students who were neither taught by the help of books in the classroom nor with the help of physical models instead with the help of any kind of software and online examination and the students communicated with the instructor on mail gave the best result in mark of the students. Overall, the findings show that teaching without the use of books gives better results and thus enhances the performance of the students. The instructor needs to develop a learning community using the E-learning pedagogy . Although this research represents a step forward in the development and evaluation of E-learning, it also raises many additional questions. How can E-learning be combined with the traditional pedagogy? More research is also needed to determine whether undergraduate and graduate business students enrolled in various courses differ in their learning needs. Research is also needed that goes beyond student perceptions to examine more quantifiable learning outcomes. Lastly, because this study was conducted at a particular institution, the generalization of these findings to others is unknown. Future research should target multiple institutions, both national and abroad. In conclusion, not only will faculty members get better E-teaching opportunity, students will also become better learners as they gain more experience with this educational medium. The end result will be improved performance of the students and overall satisfaction of the faculty and institution. Business students will come to expect highly integrated, effective, and efficient learning experiences. Those schools unwilling to commit significant resources to the endeavor will not be competitive over time.